\documentclass[fleqn,usenatbib]{mnras}
\usepackage[T1]{fontenc}
\DeclareRobustCommand{\VAN}[3]{#2}
\let\VANthebibliography\thebibliography
\def\thebibliography{\DeclareRobustCommand{\VAN}[3]{##3}\VANthebibliography}

\usepackage{graphicx}	
\usepackage{amsmath}	
\usepackage{amssymb}	
\usepackage{tikz}

\usepackage{xcolor}
\usepackage{enumitem}
\usepackage{newtxtext,newtxmath}

\title[Very metal-poor and old stars in discs of MW analogues]{On the likelihoods of finding very metal-poor (and old) stars in the Milky Way's disc, bulge, and halo} 

\author[Sotillo-Ramos et al.]{Diego Sotillo-Ramos,$^{1}$\thanks{E-mail:sotillo@mpia.de}
Maria Bergemann,$^{1}$
Jennifer K.S. Friske,$^{2}$
Annalisa Pillepich,$^{1}$
\\
$^{1}$Max-Planck-Institut f{\"u}r Astronomie, K{\"o}nigstuhl 17, 69117 Heidelberg, Germany\\
$^{2}$Mullard Space Science Laboratory, University College London, Holmbury St. Mary, Dorking, Surrey, RH5 6NT, UK
}
\begin{document}
\label{firstpage}
\pagerange{\pageref{firstpage}--\pageref{lastpage}}
\maketitle

\begin{abstract}

Recent observational studies have uncovered a small number of very metal-poor stars with cold kinematics in the Galactic disc and bulge. However, their origins remain enigmatic. We select a total of 138 Milky Way (MW) analogs from the  TNG50 cosmological simulation based on their $z=0$ properties: disky morphology, stellar mass, and local environment. In order to make more predictive statements for the MW, we further limit the spatial volume coverage of stellar populations in galaxies to that targeted by the upcoming 4MOST high-resolution survey of the Galactic disc and bulge. We find that across all galaxies, $\sim$20 per cent of very metal-poor (${\rm [Fe/H]} < -2$) stars belong to the disk, with some analogs reaching 30 per cent. About 50$\pm$10 per cent of the VMP disc stars are, on average, older than 12.5 Gyr and $\sim$70$\pm$10 per cent come from accreted satellites. A large fraction of the VMP stars belong to the halo ($\sim$70) and have a median age of 12 Gyr. Our results with the TNG50 cosmological simulation confirm earlier findings with simulations of fewer individual galaxies, and suggest that the stellar disc of the Milky Way is very likely to host significant amounts of very- and extremely-metal-poor stars that, although mostly of ex situ origin, can also form in situ, reinforcing the idea of the existence of a primordial Galactic disc.

\end{abstract}

\begin{keywords}
galaxies: spiral -- galaxies: interactions -- galaxies: structure -- Galaxy: disc -- Galaxy: structure -- Galaxy: evolution -- methods: numerical
\end{keywords}
\section{Introduction}
\label{sec:intro}

One of the most exciting questions in modern observational astrophysics is the existence of primordial, so-called Population III stars \citep[e.g.][]{Beers1985, Beers2000, Christlieb2002, Sneden2003, Frebel2005, Frebel2015}. Despite decades of theoretical and observational research, no such objects have been discovered yet, although their successors have been identified and it has become possible to link their properties to enrichment in individual stellar explosions \citep[e.g.][]{Keller2014, Howes2015, Takahashi2018, Ji2020, Yong2020, Hansen2020,Skuladottir2021, Lagaee2023}.

In this work, we explore the possibility that the Milky Way galaxy may host a primordial disc, that is, qualitatively, an old in situ disc formed at $z \gtrsim 2$ out of stars born from the rotationally-supported pristine gas. In other words, the question is whether  very metal-poor stars\footnote{Following \citet{Beers2005}, stars are defined as \textit{metal-poor} (${\rm [Fe/H]}$ $< -1$), \textit{very metal-poor} (${\rm [Fe/H]} < -2$), \textit{extremely metal-poor} (${\rm [Fe/H]} < -3$), \textit{ultra metal-poor} (${\rm [Fe/H]} < -4$), \textit{hyper metal-poor} (${\rm [Fe/H]} < -5$), and currently these categories are being extended to ${\rm [Fe/H]} < -10$.} could be hiding in the Galactic disc, in addition to their established association with the Galactic halo \citep[e.g.][]{Schorck2009, Youakim2020, Bonifacio2021} and the bulge \citep[e.g.][]{Schlaufman2014, Howes2016, Reggiani2020}. Owing to the overall progression of cosmic chemical enrichment, one expects that more metal-poor stars form in larger systems at earlier times or in smaller systems at later times and their presence would be intricately linked to the hierarchical growth of galaxies \citep{WhiteSpringel2000}. From a theoretical point of view \citep[e.g.][]{SearleZinn1978, Salvadori2010} and also confirmed by recent cosmological simulations, such as FIRE \citep{Hopkins2014}, APOSTLE \citep{Sawala2016, Fattahi2016}, and TNG50 \citep{Nelson2019b, Pillepich2019}, it is expected that most metal-poor (MP) stellar populations follow isotropic distributions of orbits and are therefore preferentially confined to the spheroidal components, bulge and halo \citep[e.g.][for TNG50]{Chen2023}. This is expected, because they form either ex situ and have been  accreted by progressively stripping smaller satellites to form mostly the halo; or they form in situ, at the early stages, when there was no rotationally supported component or in the primordial disc on orbits then heated by mergers \citep{Starkenburg2017, ElBadry2018, Chen2023}.

\begin{figure}
\includegraphics[width=\columnwidth]{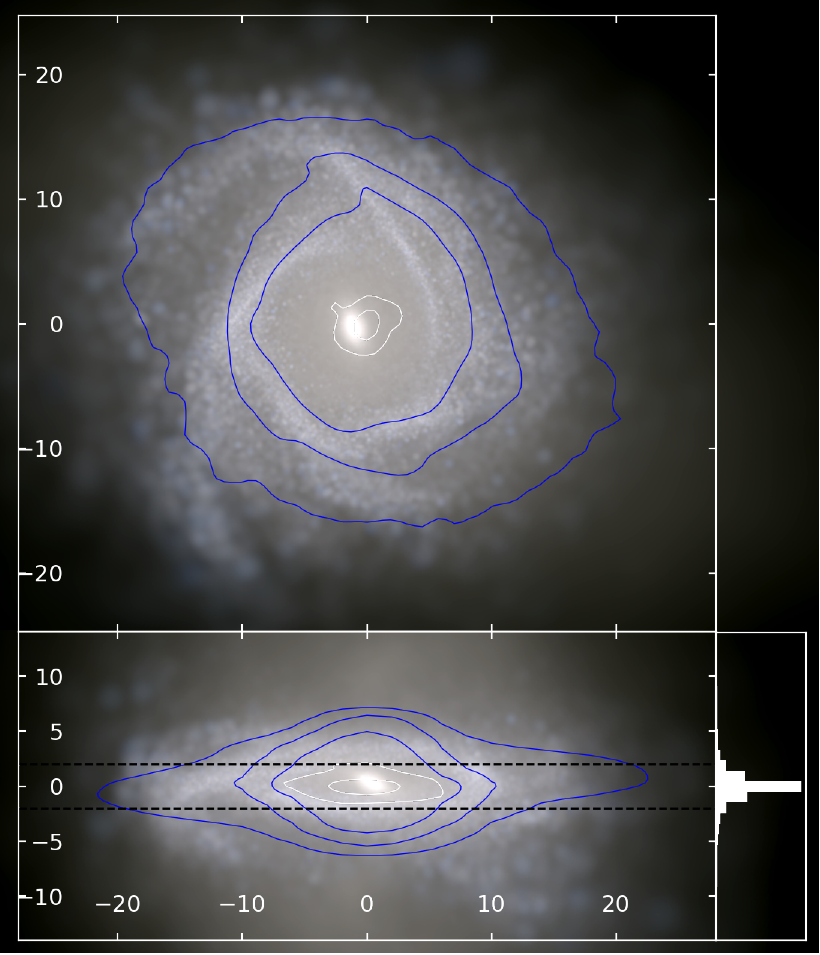}
\caption{Stellar-light composite image of one MW-like galaxy from the TNG50 simulation in face-on and edge-on projections, among 138 MW analogs. Spatial scales are given in the units of kpc. For this example (Subhalo IDs at $z=0$: 535774) disc scale-length, and think and thick disc scale-height are compatible with the current estimations for the Galaxy \citep[e.g.][]{Bland-Hawthorn2016}. Blue contours trace the stellar surface density. White contours trace the surface density of VMP stars. The side histogram shows the vertical distribution of VMP stars. In all galaxies in the sample, most of the VMP stars are concentrated very close to the midplane.}
\label{fig:1} 
\end{figure}

Observational searches have yielded over $10^5$ stars with metallicity [Fe/H] $\lesssim -2$ \citep{Bonifacio2000, Li2018, Chiti2021, Huang2022,Andrae2023}, representing  all morphological components of the Galaxy, including the bulge \citep{Howes2015, Koch2016, Reggiani2020}, the halo \citep{Hayes2018, Limberg2021}, and the disc \citep{DiMatteo2020, Sestito2020, Carter2021, FernandezAlvar2021}. With estimated fractions of $25-30$ per cent  \citep{Sestito2019,Sestito2020}, very metal-poor (VMP) stars with disky orbits are, perhaps surprisingly, not rare and are confined close to the Galaxy's midplane \citep[see also][]{Venn2020,DiMatteo2020}. There is evidence for these systems being preferentially on prograde orbits \citep{Carter2021,Carollo2023}. However, the origin of the kinematic asymmetry is currently debated.  \citet{Santistevan2021}, using FIRE-2 simulations \citep{Hopkins2018}, confirm the preference for prograde orbits for the UMP disky stars and a prograde-to-retrograde ratio of $\sim 2:1$, associating the rotational bias with a single major merger event. \citet{Sestito2021} use 5 Milky Way (MW) analogs from the NIHAO-UHD project \citep{Buck2020} to show that ${\rm [Fe/H]} < -2.5$ stars in retrograde disc orbits were accreted in the first billion years of the galaxy formation, whereas the prograde subpopulation was mostly accreted at later stages.

In this work, we use the TNG50 cosmological simulation, which is the highest resolution run of the IllustrisTNG project \citep{Naiman2018, Marinacci2018, Pillepich2018b, Nelson2018, Springel2018}, to assess the fraction of metal-poor stars expected in the different Galactic morphological components. We follow up on the analysis by \citet{Chen2023}, who performed the analysis of extremely metal-poor stars in TNG50 MW and M31-like galaxies. Differently from the latter study, here we aim to a) quantify the presence and origin of stars across a wider range of metallicity levels; b) put a special focus on the Galactic disc and c) estimate the statistics and properties of VMP, EMP and UMP stars by including their [Mg/Fe] ratios, ages and origin (in or ex situ). Crucially, we show where these stars are distributed in MW simulated analogues by using the nominal spatial selection informed by the upcoming 4MOST high-resolution disc and bulge survey \citep{Bensby2019}, in order to provide predictions for the detectability of VMP stars in next-generation observational programs. Compared to previous works based on zoom-in simulations, we increase the MW-analogs sample size by a factor of $\sim$10-20.

The paper is organised as follows: in Sec.~\ref{sec:methods} we describe the cosmological simulation TNG50 and how the MW analogs are selected. We describe in Sec.~\ref{sec:MPstars} the populations of metal-poor stars across morphological components, exploring additional properties such as their ages, [Mg/Fe] abundances and origin. We discuss possible implications for the current understanding of the origin and composition of the Galaxy's disc in Sec.~\ref{sec:conclusions}.
\section{Methods} \label{sec:methods}
In this paper, we focus on MW-like galaxies realized within the cosmological simulation TNG50 \citep{Nelson2019b, Pillepich2019}. For details on the simulation, we refer to the latter papers, and here provide a brief account of the main properties of the galaxies.

The simulation comprises a periodic cubic volume with a side length of $51.7$ comoving Mpc, contains 2160$^3$ dark matter (DM) particles and equal initial number of gas cells. The DM particles have an uniform mass of $m_\rmn{DM} = 4.5 \times 10^5$ M$_{\odot}$, while the gas cells (and stellar particles) have an average (initial) mass of  is $m_\rmn{baryon} = 8.5 \times 10^4$ M$_{\odot}$. The star formation follows \citet{SpringelHernquist2003}: the gas is transformed into star particles stochastically when the density exceeds $n_{\rm H} = 0.1{\rm cm}^{-3}$ on time scales to reproduce the Kennicutt-Schmidt relation \citep{Kennicutt1989}. Stellar particles represent stellar populations that are born at the same time and have an initial mass distribution by \citet{Chabrier2003}. Detailed information about all the particles, subhaloes, and haloes, is stored in 100 snapshots. The (sub)haloes at different snapshots, i.e. across cosmic times, are linked via the \textsc{Sublink} \citep{RodGom2015} and \textsc{LHaloTree} \citep{Springel2005a} algorithms, so that the assembly histories of galaxies is available. In this paper, we use the baryonic version of \textsc{Sublink} and the main progenitor of a galaxy is the one with the most massive history. We will also identify \textit{in situ} stars as the ones formed in the main progenitor; accreted stars will be referred to as \textit{ex situ} \citep[as per ][]{RodGom2016}.

From the TNG50 simulation box, which returns hundreds of massive galaxies at $z=0$, \citet{Pillepich2023} identify the 198 most suitable counterparts to the MW and M31 based on their properties at $z=0$ according to galaxy stellar mass, stellar diskyness, and environment. This galaxy sample has been previously used and extensively detailed in terms of its stellar content also by \citet{Engler2021, SotilloRamos2022, Engler2023, Chen2023}. With an additional cut in galaxy stellar mass ($10^{10.5-10.9}\,{\rm M_{\odot}}$), in this paper we identify the 138 best MW analogs from TNG50 at $z=0$. We note that this selection does not impose any constraints on the evolutionary paths of galaxies nor a-priori on the detailed structural and chemical properties of the stellar disc and bulge.

In Fig.~\ref{fig:1} we show the stellar-light composite face-on and edge-on images of one MW-like galaxy from the TNG50 sample, at $z=0$. This simulated analog has disc scale-length and thin and thick disc scale-heights compatible with the current estimations for the Galaxy \citep[see][for more details on the calculations and the MW reference values]{SotilloRamos2023}. We overlay the positions of VMP stellar particles with white contours and shows that a high fraction lies within a few kpc of the midplane, as is the case for the Galaxy.
\subsection{Morphological decomposition of MW analogues}
\label{sec:morpho}

There are many methods to decompose a galaxy in its stellar morphological components. Recent methodologies applied to simulated galaxies have been described by \citet{Du2019, Gargiulo2022, Zhu2022}, are based on earlier works by e.g. \citealt{Abadi2003, DomenechMoral2012, Obreja2018} and combine structural and kinematical information. In this paper, we choose the approach by \citet{Zhu2022}, which is based on the orbit circularity $\epsilon_{\rm z}$\footnote{$\epsilon_{\rm z}=j_\rmn{z}/j_\rmn{circ}$, with $j_\rmn{z}$ the specific angular momentum of the star in the direction perpendicular to the galactic disk, and $j_\rmn{circ}$ the specific angular momentum of a star at the same radius, on a circular orbit, i.e., $j_\rmn{circ}=rv_\rmn{circ}$, with $v_\rmn{circ}=\sqrt{{\rm G}M(\leq r)/r}$ the circular velocity of the galaxy at the considered radius.}
\citep[as defined by][]{SotilloRamos2022} and galactocentric distance $r_{\ast}$ of the stars. In brief, the four main stellar components are defined as follows \citep[see also][]{Chen2023}:
\begin{itemize}
    \item Cold disk: $\epsilon_{\rm z} > 0.7$
    \item Warm disk: $0.5<\epsilon_{\rm z}<0.7$ and $r_{\rm cut}<r_{\ast}<r_{\rm disk}$
    \item Bulge: $\epsilon_{\rm z}<0.7$ and $r_{\ast}<r_{\rm cut}$
    \item Stellar halo: $\epsilon_{\rm z}>0.5$ and $r_{\rm disk}<r_{\ast}<r_{\rm halo}$ or $\epsilon_{\rm z}<0.5$ and $r_{\rm cut}<r_{\ast}<r_{\rm halo}$,
\end{itemize}
with $r_{\rm cut}=3.5~{\rm kpc}$, $r_{\rm disk}=6\times r_{\rm d}$, the exponential scale-length of the stellar disk, as measured by \citet{SotilloRamos2022}, $r_{\rm halo}=300~{\rm kpc}$ the maximum galactocentric distance to which we consider that the stellar halo extends: see also fig. A1 in \citet{Chen2023} for a visual depiction of the components in the $\epsilon_{\rm z}-r_{\ast}$ plane. 'Cold' and 'warm' discs are similar, but are {\it not} apriori equivalent, to the geometrically defined 'thin' and 'thick' discs based on fitting the vertical stellar density profiles \citep[e.g.][]{GilmoreReid1983}.
\begin{figure}
\includegraphics[width=1\columnwidth]{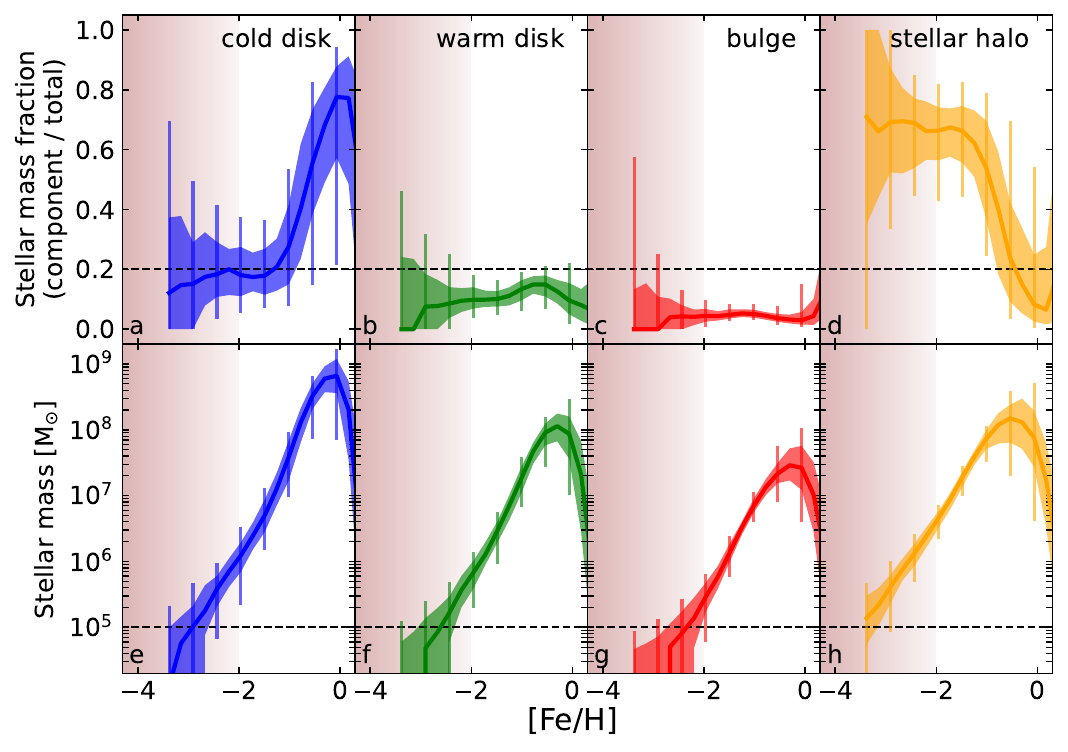}
\caption{Metallicity distributions of stars in TNG50 MW-like galaxies, grouped by their respective morphological component: cold disc, warm disc, bulge and stellar halo, from left to right. The top panels quantify the stellar mass fractions in each component to the total stellar mass; the bottom ones are the metallicity distributions in each morphological component, in stellar mass. The solid lines represent the medians and the shaded areas and error-bars represent inter-per centile ranges across the galaxy sample: 16 to 84 and 2 to 98, respectively. The vertical red shaded bands highlight metallicities [Fe/H]$\leq -2$, i.e., VMP stars. We remind the reader that the distributions are shown for the characteristic observable volume fraction of the Galaxy, as it will be `seen' by the 4MIDABLE-HR survey on 4MOST (see Section~\ref{sec:morpho}). Specifically, we apply a volume cut of $5.5$ kpc in heliocentric distance, where the fiducial 'Sun' is placed at $8$ kpc in the simulated galaxy. }
\label{fig:2}
\end{figure}

In the majority of the TNG50 MW analogues (96), the cold disc is the most massive component (by galaxy selection), with median values of $\sim 1-3 \times 10^{10}$ M$_{\odot}$, and increasing with galaxy stellar mass. There is also a significant, more than half an order of magnitude, galaxy-to-galaxy variation in the mass of all morphological components. The warm disc is the least massive component and it is one order of magnitude less massive than the cold disk. The bulge is in most cases the second most massive component, except for some galaxies at the high-mass end, where the stellar halo is more massive. 

In relative terms, the cold disc represents $\sim$60 per cent of the total stellar mass for the less massive MW analogs and decreases to $\sim$40 per cent in the case of the most massive ones in the sample. The relative contribution of the other components does not change significantly with galaxy stellar mass. This dominance of the disc is related to the selection of the galaxies: it is in good agreement with the analysis of tens of edge-on spiral galaxies in the local universe by \citet{Comeron2014} and, considering the D/T fraction, with the properties of MW analogs from the NIHAO simulations by \citet{Obreja2018}. 

In this paper, we focus on the qualitative comparison with the expected properties of stars to be observed within the 4MIDABLE-HR survey \citep{Bensby2019} that will be carried out at the 4MOST facility \citep{deJong2019}. This survey will provide coherent homogeneous characterisation of a very large number (over $3$ million) of stars in the Galactic disc and bulge, including their detailed metallicities, abundances, ages, and kinematics. To mimic the spatial coverage of 4MIDABLE-HR, we apply a cut on the volume occupied by the simulated stellar particles of 5.5 kpc in heliocentric distance centered at a random point positioned at 8 kpc from the galaxy center. We note that the 4MIDABLE-HR survey selects the targets based on apparent magnitude but, owing to instrumental limits and the complex observational strategy, in practice most stars to be observed (with the exception of selected fields) will be confined within the given spatial volume: this lends credibility to our procedure, despite its simplicity. It should also be noted that, with this spatial cut, the bulge component is represented only by its most external stars. Finally, we have checked that the results of the paper are qualitatively the same than if we had placed the fiducial 'Sun' not at a fixed $8$ kpc distance but at 4 times the disc length of each galaxy, to account for the diversity in galaxy sizes \citep[see e.g. Figure 13 of][]{Pillepich2023}.

In the next section, we explore in detail the temporal, chemical, and evolutionary properties of these four main Galactic components, and analyse their distributions by focusing on the metal-poor and old populations.
\section{Results} \label{sec:MPstars}
We begin with the analysis of the temporal (i.e. of the stellar ages) and chemical properties of the components in the simulated galaxies, and then proceed with the assessment of the statistical properties of the distributions in the volume that will be accessible to next-generation spectroscopic surveys of the Galaxy, such as with 4MOST \citep{Bensby2019, Chiappini2019, Christlieb2019}, WEAVE \citep{Dalton2012, Dalton2016}, and MOONS \citep{Gonzalez2020}. As described in the previous section, the focus here is on the 4MOST 4MIDABLE-HR survey and we refer the reader to the science case for more details on its scope and strategy \citep{Bensby2019}.
%
%
%
%
\subsection{Trends with metallicity}\label{sec:res1}
Fig.~\ref{fig:2} shows the metallicity distribution functions (MDFs) for all four Galactic components of all TNG50 MW-like galaxies described in Section~\ref{sec:methods}. The distributions include a cut on the volume to mimic the spatial coverage of 4MIDABLE-HR, as detailed in the previous Section. The solid lines represent the medians across galaxies. Shaded areas and error-bars represent inter-per centile ranges also across the studied galaxy sample: 16 to 84 and 2 to 98 per cent, respectively. Cold disk, warm disk, bulge and halo are represented, respectively, with blue, green, red and orange lines, from left to right. We also show the stellar mass fraction per component, that is stars in the component relative to the total number of stars: top row. 

In most TNG50 MW analogues, the majority of Sun-like, [Fe/H]$\sim 0$ stars reside in the kinematically cold 'thin' disk, but with some scatter that encompasses $\sim 20$ per cent of the total number of stars in the disc. The bulge is the second most populated component when no cut is applied, whereby we note that the apparent difference in the total stellar mass in the bulge and in the thin disc is caused by the application of the fiducial volume cut to account for the survey selection of 4MIDABLE-HR. In general, just a small, although non-negligible (at the level of 20 per cent) fraction of the solar-metallicity stars can be found in the halo or the warm disk, although we note a significant galaxy-to-galaxy variation. 

For stars with [Fe/H] $\lesssim -1$ the trend reverses. As expected, in the vast majority of galaxies, most of these low-metallicity stars reside in the stellar halo, with a median fraction of $\sim~60$ per cent. However, and this is one of most interesting results, the fraction of VMP stars in the cold disc component still reaches up to $\sim~20$ per cent, in  the typical galaxy. In fact, in some MW-like galaxies, as many as 40 per cent of the stars with [Fe/H]$\sim -2$ follow cold disky orbits. For progressively lower metallicity values, the trends change slope. Most of the stars with metallicity values [Fe/H] $\lesssim -2$ across all galaxies can be found in the stellar halo, with the median fraction of $\sim 80$ per cent. For the cold (thin) disk, the median values are around $\sim 15$ per cent, but we consistently find a significant ($\sim$25 per cent) number of galaxies where the fraction of VMP stars is $\geq 25$ per cent. These results are qualitatively and quantitatively very similar if we place the 'Sun' at four times the disc length: the median values change only minimally across metallicity and component, although the scatter is in all cases is larger.

In summary, we find that large fractions of ultra, extremely and very metal-poor stars are present in all morphological components of the simulated Milky Way analogues. Most intriguingly, their MDFs suggest that such stars should be abundantly present in the cold disk, typically referred to as the thin disc.
This finding confirms recent results of kinematically cold VMP targets in the literature \citep{Sestito2019, DiMatteo2020}, with our results indicating that the fraction of such stars in the Galactic disc could be even higher than the current observational evidence suggests. 
\begin{figure}
\includegraphics[width=\columnwidth]{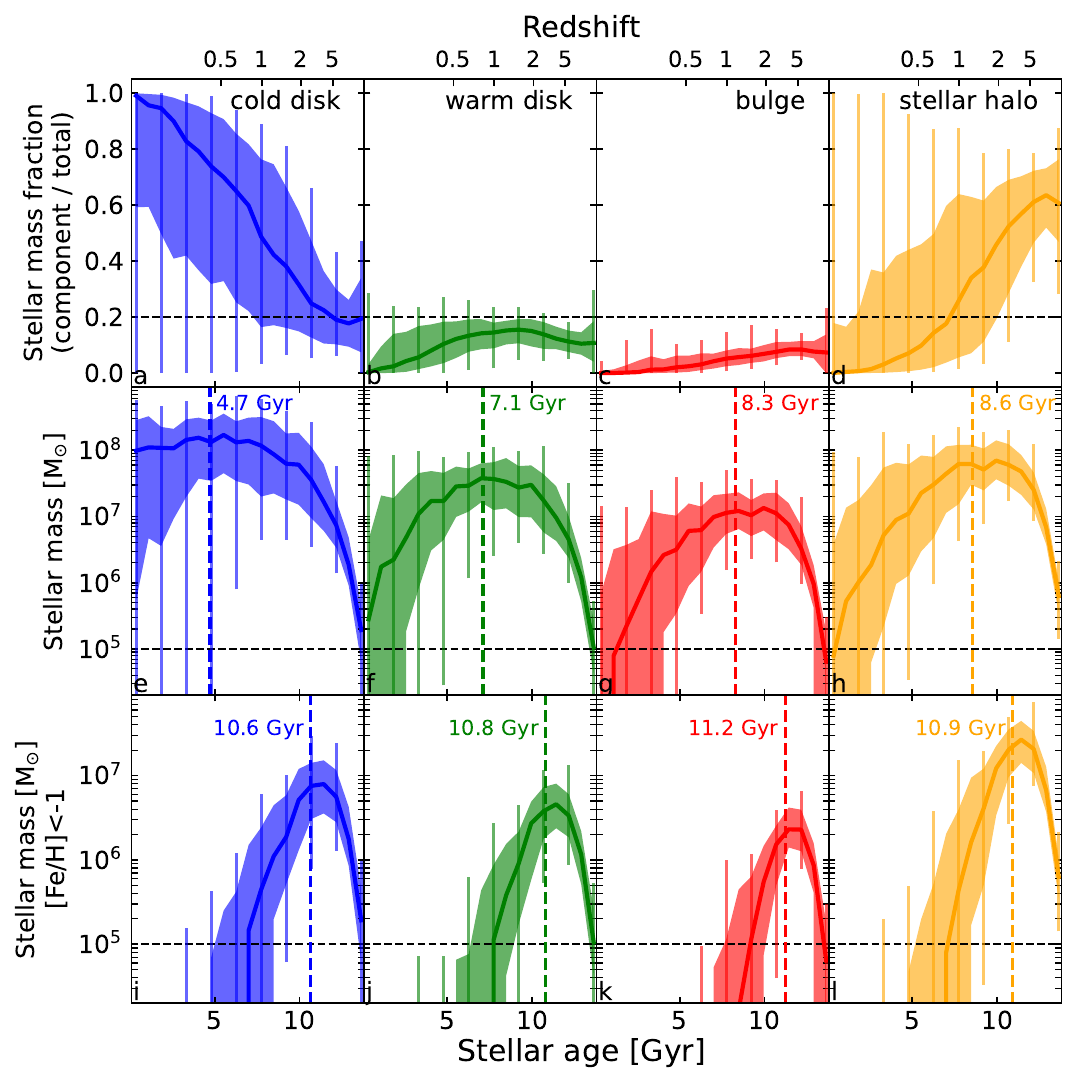}
\caption{Same as Fig. 2, but for the distribution of stellar ages in the morphological components. \textit{Top}: Stellar mass fraction per component to total stellar mass. \textit{Middle}: Stellar mass per component. \textit{Bottom}: Mass of MP stars per component. The vertical dashed lines represent the median age of the stars in the component (within the volumetric cut), across all galaxies.} 
\label{fig:3} 
\end{figure} 
\subsection{Trends with stellar age}\label{sec:res2}
In order to understand the temporal history of metal-poor populations, in Fig.~\ref{fig:3} we show the age distribution functions of the halo, bulge and disc components in the simulated galaxies. In the top row, we normalise the fraction of stars per [Fe/H] bin in each population to the total number of stars in all populations. In the second row, as in Fig.~\ref{fig:2}, we also provide the stellar mass in each component per metallicity bin, and in the bottom row we show the corresponding mass of each stellar population for metal-poor stars with [Fe/H] $< -1$.  

The bulge and stellar halo appear as the oldest components, followed by the warm disc and the cold disc as the youngest component, although each of these populations show a significant temporal extent spanning the entire range of ages up to $\sim 13$ Gyr, with only a mild dependence on the galaxy. Even in the cold disc, a non-negligible fraction of stars of  $\sim 20$ per cent, have ages greater than 10 Gyr, and the cold discs of some MW analogues stand out with fractions as large as $50$ per cent. There is no galaxy in our TNG50 sample that does not host an old cold disc. On the one hand, properties of these distributions are consistent with what observers would usually describe as the canonical formation picture of the MW \citep{Freeman2002}. On the other hand, the extended star formation histories of all components, especially that of the disc and (to a lesser extent) of the bulge, are striking and indicate that the Milky Way may host a primordial disc that we explore in more detail in Sec.~\ref{sec:origins}.

\begin{figure}
\includegraphics[width=1\columnwidth]{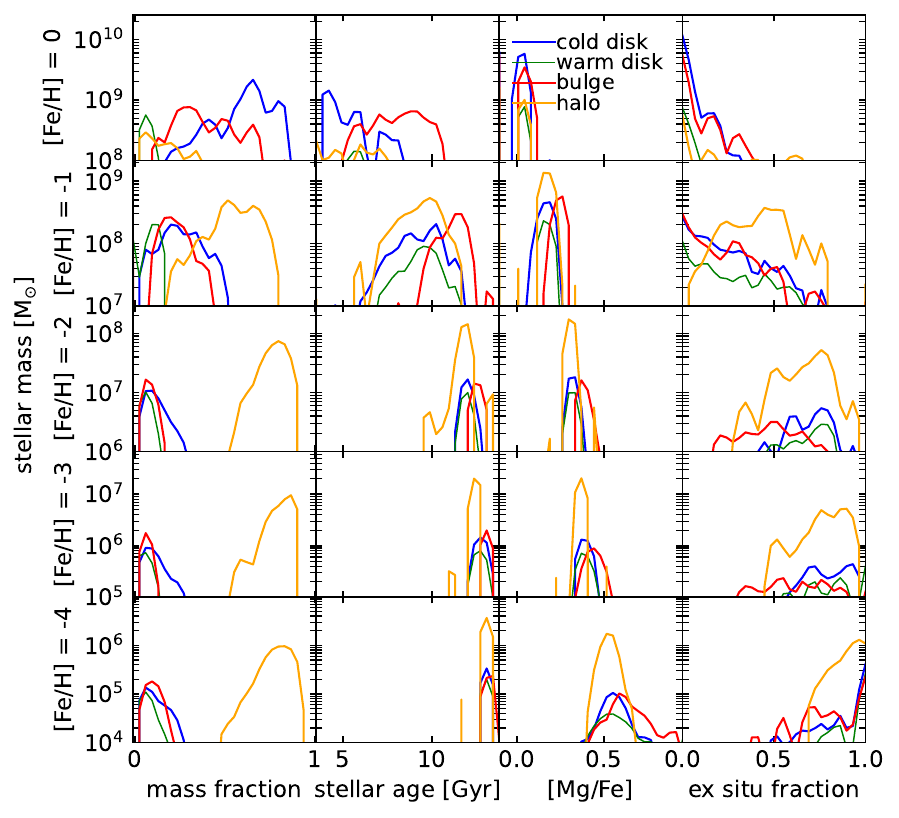}
\caption{\label{fig:histograms} Mass fraction per component, stellar age, [Mg/Fe] and ex situ fraction distributions, for stellar samples with different values of [Fe/H]. We show the distributions of the median values across all TNG50 138 MW-like galaxies weighted by stellar mass and do not apply a heliocentric cut.}
\end{figure}

In the bottom row of Fig.~\ref{fig:3}, we show the age distributions for the metal-poor populations ([Fe/H] $<-1$) of the discs, the bulge, and the stellar halo. Here we do not apply the heliocentric cut, in order to be able to find a significant enough number of EMP and UMP stars. It is clear, and expected, that the age distribution of each component exhibits a trend toward older ages with decreasing metallicity: middle vs. bottom panels of Fig.~\ref{fig:3} \citep{Bergemann2014}. Specifically, the median age of all four components is now skewed towards ages $\geq 8$ Gyr, and the mode values peak at $\gtrsim 12$ Gyr for all galaxies and components. Cold disc stars with [Fe/H]$\approx -1$ have an age distribution that closely resembles that of the warm disc and the halo (with the median ages of $\sim$10 Gyr to 11 Gyr), whereby the bulge appears to be made of the oldest population, as consistently seen in all MW analogues. We find very similar distributions for VMP, EMP and UMP stars. We also note, and this will be discussed in more detail in the next section, that generally, halo stars of all metallicities are on average younger than the stars in the bulge (Fig.~\ref{fig:histograms}). 

Finally, by exploring the median [Mg/Fe] abundance ratios (Fig.~\ref{fig:histograms}), we also find a strong dependence of the distributions on metallicity. In line with observational evidence \citep[e.g.][]{Bensby2014, Bergemann2017, Nissen2018}, [Mg/Fe] increases as the [Fe/H] values decrease (from the top to the bottom panels). At [Fe/H]$\sim 0$, the cold disc has the lowest [Mg/Fe]. For metal-poor stars the bulge exhibits, on average, slightly higher (by $\sim 0.05$ dex) values of [Mg/Fe]. Such a systematic difference for the bulge is indeed observed in the Milky Way \citep{RichOriglia2005, CunhaSmith2006, Ryde2010, Rich2012}, but see also \citet{Joensson2017} and \citet{Griffith2021}, who advocate smaller chemical differences between the disc and the bulge (but with the caveat here that the different works use different definitions for the morphological components). Interestingly, we also find an increasingly large scatter of [Mg/Fe] ratios for $-4 \lesssim $ [Fe/H] $\lesssim -3$, which is consistent with the observational distributions of EMP stars by \citet{Howes2016}, although their data suggest a higher scatter also for stars with [Fe/H] $\lesssim -2.5$. At higher metallicities, [Fe/H] $\gtrsim -2$, the stellar halo and both disc components show similar trends of progressively declining [Mg/Fe] values. However, we emphasize that owing to very large differences in the stellar mass in each metallicity bin, it is unlikely that one would observe many low-$\alpha$ halo stars or high-$\alpha$ cold disc stars, respectively.
\subsection{Origins of metal-poor stars}\label{sec:origins}
The detailed properties of the stellar particles in the TNG50 simulation allow the identification of their origin, in particular, whether they formed in the main galaxy or whether they were accreted. Fig.~\ref{fig:histograms} shows that as with the age and [Mg/Fe] abundance, there are clear trends that depend on the stellar metallicity. More metal-poor stellar populations have, on average, a higher ex situ fraction. This applies to all Galactic components, although the halo, on average, has a comparatively higher ex situ fraction  for all metallicities. These distributions are qualitatively consistent with observations, e.g. \citet{Conroy2019} have shown that the accreted fraction increases with decreasing metallicity for the Milky Way halo stars. 

Kinematically cold and warm discs are dominated by the ex situ component for metallicities below [Fe/H] $\lesssim -2$. Yet, interestingly, even at the lowest metallicities a non-negligible fraction of stars in these components have an in situ origin. The median ex situ fraction values for the discs are, in these cases, $\approx65$, $\approx75$ and $\approx100$ per cent, for metallicities [Fe/H] of $-2$, $-3$, and $-4$, respectively. However,  importantly, we find also some MW analogs (Fig.~\ref{fig:histograms} bottom row, right) where $10$ to $30$ per cent of the UMP disk stars were formed in situ, and given the age of these stars, most likely in a primordial disk.
This finding is not in contradiction with \citet{Ruchti2015}. They report no strong evidence for an accreted disc component, however, their observed sample is dominated by targets with [Fe/H] $\gtrsim -1$, and only a very small fraction of their stars are more metal-poor than [Fe/H] $\sim -2$, which, as we see in Fig.~\ref{fig:histograms}, represents a transition from the in situ to the ex situ dominated regime. Indeed, TNG50 suggests that most of the stars with [Fe/H]$\gtrsim -1$ have a strongly in situ dominated origin, regardless of their parent Galactic component. However, for some of the analogs ($\sim$10 per cent), the accreted disc fraction can be also significant ($\gtrsim$15 per cent).
The properties of early galactic discs, such as their alignment with respect to the present-day orientation,  are complex and will be a subject of another paper. This has also been addressed, e.g. in \citet{BelokurovKravtsov2022}. However, here we note that from a quick inspection of the angular momenta, we find that at ancient times, galaxies show a large spread of angular momentum vectors  close to a uniform distribution. This could be expected if stars formed in a chaotic way, with multiple gas inflows from many random directions and mergers, that potentially destroy and heat the primordial discs and bring stars in randomly distributed orbits. We cannot detect any preferred angle for the orientation of the primordial disk, but the alignment steadily increases as the galaxies evolve  and approach the present-day age, $z=0$.
\section{Summary and conclusions}\label{sec:conclusions}
We have used the cosmological magneto-hydrodynamic galaxy simulation TNG50 to explore the fraction of very metal-poor stars, [Fe/H] $\lesssim -2$, in Milky Way like galaxies. The selection of galaxies follows our detailed previous work presented in \citet{Pillepich2023}. We furthermore apply observationally motivated limits to the theoretical distributions,  aiming to understand which fraction of the metal-poor stars would be observable in the Galactic disc, bulge, and halo with next-generation facilities, such as the 4MIDABLE-HR survey on 4MOST \citep{Bensby2019}.

Through the statistical analysis of the stellar populations of simulated galaxies, specifically their metallicities, ages, and Mg/Fe ratios, we find that metal-poor stars are common in all morphological components of the MW analogues. As expected, the  stellar halo is the component primarily hosting VMP stars \citep[see also][]{Chen2023}. However, we find that in the cold `thin' discs of TNG50 MW-like galaxies, the fraction of VMP, EMP, and UMP stars is typically $\approx 20$ per cent of the total number of stars, and in some MW-like galaxies stars with [Fe/H] $\approx -3$ reach up to 50 per cent. Most of these low-metallicity stars are formed ex situ. The temporal properties of these populations suggest that all galaxy components, i.e. the cold (thin) disc, the warm (thick) disc, the halo, and the bulge, have very extended evolutionary histories with ages reaching $\gtrsim13$ Gyr. This suggests that the Galaxy could host a primordial cold disc even though a significant fraction of metal-poor stars in disc orbits originated from  small satellites and got subsequently accreted. 

The large unbiased sample of MW analogs of the TNG50 simulation confirms that metal-poor stars can help unveil the first steps of the formation of the Galaxy. Contrary to what has been largely expected thus far, it is very likely that many of these stars follow cold disc orbits. 
We hence recommend that current and future Galactic surveys should target not only the stellar halo, but also the disc for the search of the most metal-poor old stars, which will be equivalent to exploring the regime of redshifts $z\sim 2$ to $>5$.

\section*{Acknowledgements}
We are grateful to the anonymous referee for the insightful suggestions that helped to improve the work.
We acknowledge Timothy C. Beers and Anirudh Chiti for valuable discussions.
DSR, AP, and MB acknowledge support by the Deutsche Forschungsgemeinschaft (DFG, German Research Foundation) -- Project-ID 138713538 -- SFB 881 ("The Milky Way System", subprojects A01, A05, A06, A10). MB is supported through the Lise Meitner grant from the Max Planck Society. This project has received funding from the European Research Council (ERC) under the European Unions Horizon 2020 research and innovation programme (Grant agreement No. 949173).
JF acknowledges support from University College London’s Graduate Research Scholarships and the MPIA visitor programme.
The TNG50 simulation was realized with compute time granted by the Gauss Centre for Super- computing (GCS), under the GCS Large-Scale Project GCS-DWAR (2016; PIs Nelson/Pillepich) on the GCS share of the supercomputer Hazel Hen at the High Performance Computing Center Stuttgart (HLRS). 
This work benefited from a workshop supported by the National Science Foundation under Grant No. OISE-1927130 (IReNA), the Kavli Institute for Cosmological Physics, and the University of Chicago Data Science Institute.

\section*{Data Availability}

Data directly related to this publication and its figures are available on request from the corresponding author. The IllustrisTNG simulations, including TNG50, are publicly available and accessible at \url{www.tng-project.org/data} \citep{Nelson2019a}. A special data release is also available for the TNG50 Milky Way and Andromeda like galaxies, as per \citet{Pillepich2023}.


\DeclareRobustCommand{\VAN}[3]{#3}
\bibliographystyle{mnras}
\bibliography{MWM31_biblio}



\label{lastpage}

\end{document}